# Hydrogenation of single-walled carbon nanotubes

A. Nikitin[1], H. Ogasawara[1], D. Mann[2], R. Denecke[1)*], Z. Zhang[3], H. Dai[2], KJ Cho[3], A. Nilsson[1,4]

[1] Stanford Synchrotron Radiation Laboratory, 2575 Sand Hill Road, Menlo Park, CA 94025, USA
[2] Department of Chemistry, Stanford University, Stanford, California 94305
[3] Department of Mechanical Engineering, Stanford University, Stanford, California 94305
[4] FYSIKUM, Stockholm University, Albanova University Center, S-10691 Stockholm, Sweden

Towards the development of a useful mechanism for hydrogen storage, we have studied the hydrogenation of single-walled carbon nanotubes with atomic hydrogen using core-level photoelectron spectroscopy and x-ray absorption spectroscopy. We find that atomic hydrogen creates C-H bonds with the carbon atoms in the nanotube walls and such C-H bonds can be completely broken by heating to 600 $^{o}$C. We demonstrate approximately $65 \pm 15$ at % hydrogenation of carbon atoms in the single-walled carbon nanotubes which is equivalent to $5.1 \pm 1.2$ weight % hydrogen capacity. We also show that the hydrogenation is a reversible process.

*Permanent address: Lehrstuhl für Physikalische Chemie II, Universität Erlangen-Nürnberg, Egerlandstr. 3, D-91058 Erlangen, Germany

Safe, efficient and compact hydrogen storage is a major challenge for realizing hydrogen powered transport. Thus, a media that absorbs and releases a large quantity of hydrogen easily and reliably is being actively sought. Since Dillon et al [1] showed that carbon nanotubes can store hydrogen, this material has been considered as a candidate for hydrogen storage media.

Previous studies have reported hydrogen capacities for carbon nanotubes ranging from 0.25 to 20 weight % [2-4]. The reasons for such large variation in experimental results can be interpreted in the following ways. Nanotube samples often contain impurities that can influence the hydrogen adsorption (amorphous carbon, water, hydrocarbons), and the amount of impurities can change considerably between different studies in an uncontrollable manner leading to large systematic errors [2,3]. Also, different samples have varying diameter distributions, which can strongly affect hydrogen uptake performance [5-7]. Consequently, it is essential to determine the hydrogen uptake on well defined samples.

Physisorption and chemisorption both have been proposed as possible mechanisms for hydrogen storage in carbon nanotubes and according to the theoretical calculations the strength of the interaction between physisorption and chemisorption can vary between 0.11 eV per $H_2$ molecule [7] and ~2.5 eV per H atom [5,7,8]. The physisorption mechanism involves the condensation of molecular hydrogen inside or between the nanotubes. [1-3,6,7]. The chemisorption mechanism requires dissociation of hydrogen using a catalyst that subsequently leads to a reaction with the unsaturated C-C bonds to form C-H bonds [9]. While most previous studies have focused on the hydrogen storage through physisorption, recent Density Functional Theory (DFT) calculations for single-walled carbon nanotubes (SWCN) [5, 8, 10] indicate up to 7.5 weight % hydrogen storage through chemisorption by saturating the C-C double bonds on the nanotube walls. DFT calculations also predict a variation in the C-H bond strength depending on the nanotube diameter [5]. This provides a tuning mechanism for releasing stored hydrogen at ambient temperatures and pressures. Previously it was shown by infrared spectroscopy [11] and ultraviolet photoelectron spectroscopy [12] that atomic hydrogen can create C-H bond with carbon atoms of SWCN. However, experimental evidence directly demonstrating that a sufficient amount of hydrogen can be stored by chemisorption is still urgently needed.

To observe chemical interaction of hydrogen with carbon nanotubes it is essential to perform measurements using techniques that are sensitive to the formation of C-H bonds. It is also important to quantify the amount of hydrogen that is chemically adsorbed in terms of per carbon atom. This can be accomplished using atom specific techniques involving the C1s level from the carbon atoms in the nanotubes. In X-ray Photoelectron Spectroscopy (XPS) chemical shifts of the C1s level due to hydrogen coordination can provide both chemical identification of C-H bonds and from the relative intensity the number of affected carbon atoms [13]. X-ray absorption spectroscopy (XAS) provides means to observe the unoccupied orbital structure using a core excitation process [14]. The formation of C-H bonds can be observed through the modification of the carbon nanotube electronic structure around specific carbon atoms.

In this letter, we report hydrogenation of SWCN films with an atomic hydrogen beam [12]. This approach allowed us to exclude the influence of hydrogen dissociation on the hydrogenation mechanism. Using XAS, we demonstrated a decrease of π* character from unsaturated C-C bonds in the walls of SWCN and an increase in the C-H* resonance due to hydrogenation. On the basis of XPS results combined with theoretical calculations, we measured $65 \pm 15\%$ hydrogenation of carbon atoms in the SWCN film that corresponds to $5.1 \pm 1.2$ weight % hydrogen capacity. The formed C-H bonds are stable at ambient temperature and break completely at 600 $^{o}$C. We also show that the hydrogenation is reversible. These results clearly demonstrate that the hydrogen storage through chemisorption in SWCN films is possible with a technologically sufficient hydrogen weight capacity.

Thin SWCN films were prepared on a silicon wafer covered with a thin layer of native oxide. Catalyst for the growth was prepared by the sonication of a mixture of 50mg Degussa Aerosil 380 Silica, 4.5 mg of Cobalt Acetate and 3.1 mg of Iron Acetate in ethanol for 2 hours [15]. The catalyst was spin-coated on the surface of silicon wafer at 3000 rpm. The SWCN films were grown by flushing a 1"quartz tube containing the prepared wafer with 470 sccm of $H_2$ for 10 minutes, heating up to 850 °C in 470 sccm of $H_2$ and then at 850 °C in 400 sccm $CH_4$ / 70 sccm $H_2$ which was flowed for 7 minutes, followed by cooling in 470 sccm of $H_2$ [16]. The radial breathing mode (RBM) position in the Raman spectra of such films (see insert of fig. 1) indicated a distribution in nanotube diameters ranging from 1 to 1.8 nm [17]. A scanning electron microscopy (SEM) study also showed that the nanotubes did not cover the substrate completely and partly were found in bundles.

The experiments were performed at beamline 5-1 at Stanford Synchrotron Radiation Laboratory (SSRL) and at beamline 11.0.2 at the Advanced Light Source (ALS). The energy resolution of the XPS spectra was better than 0.1 eV and the XAS spectra were measured with a resolution of 0.1 eV. An atomic hydrogen beam was produced by thermal cracking of molecular hydrogen in a W capillary [18,19]. For XPS and XAS measurements samples were hydrogenated separately in-situ. The temperature of the SWCN film was assumed to be close to the temperature of the substrate which was measured using a type K thermocouple. The SWCN films were cleaned by carefully annealing up to 750 °C with an operating pressure remaining below $1 \times 10^{-9}$ torr. The XPS spectra showed no signals from any impurities in the SWCN films. No signal from residual metal catalyst, Fe or Co, was observed since the concentration was less than 0.1%, below the XPS detection limit. We also did not observed any influence of charging in the XPS spectra.

Raman measurements were carried out in air using a Renishaw micro raman spectrometer with an excitation wavelength of 785 nm. The G and D bands (~1610, ~1350 $cm^{-1}$, respectively) are evident in all scans, as well as RBM of the SWCN (see fig. 1, insert).

The expected C1s chemical shifts were calculated in the framework of DFT and in the transition state approximation [20] using the StoBe software package [21]. Becke 88 GGA (Generalized Gradient Approximation) exchange functional [22] and Perdew-Wang 91 GGA correlation functional [23] were used for the calculations. The structures of the pristine nanotubes and nanotubes with one adsorbed hydrogen atom were obtained from molecular dynamics optimization using the Adaptive Intermolecular Reactive Empirical Bond Order potential [24].

Plot (a) in fig. 1 shows the C1s XPS spectrum for a clean SWCN film. A narrow C1s line at 284.75 eV is observed which has a full width at half maximum (FWHM) of 0.44 eV. The observed FWHM is larger than the FWHM of C1s line for graphite (FWHM=0.32 eV) [25] but is more narrow in comparison with previous studies of SWCN [26,27]. Annealing to 750 °C did not lead to a change in the FWHM of the C1s line indicating no desorption of hydrocarbon contaminations or rearrangement of carbon species with dangling bonds such as amorphous carbon or other defects. The insert in fig. 1 shows a Raman spectrum of the clean SWCN film used in the current study. We observe that the G to D band ratio is large indicating a small number of the defects. We therefore assume that the studied SWCN films are free of any major contaminants and defects that could influence the hydrogen adsorption process.

Plot (b) in fig. 1 shows the room temperature C1s XPS spectrum of SWCN after in-situ exposure to the atomic hydrogen beam. We see that hydrogen interaction leads to a dramatic change in the C1s peak shape. There is an increase in the FWHM to 1.3 eV and an additional ("second") C1s peak is recognized as a shoulder at higher binding energy compared to the initial peak maximum. We assign the "second" peak to an energy position between 285 eV and 286 eV with a FWHM ~1 eV. Furthermore, there is a dramatic decrease of intensity at 284.75 eV corresponding to carbon atoms in pure nanotubes. The observed 0.8 eV binding energy shift between main peak and the "second" peak is in accordance with the chemical shift between $sp^2$ and $sp^3$ hybridized carbon species in nanotube materials [27]. In the present case, the result can be interpreted as a rehybridization from $sp^2$ to $sp^3$ due to attachment of hydrogen to the C-C π-bonds resulting in C-H bond formation.

Fig. 2 shows C K-edge XAS spectra for the clean (plot (a)) and the hydrogen covered SWCN films (plot (b)) which confirm the carbon atom rehybridization due to C-H bond formation. In the plot (a) two prominent spectral features are seen at 285.5 and ~293 eV, correspond-ing to the π* and σ* resonances, respectively [13,14]. This is in agreement with the general electronic structure in π-conjugated carbon materials as probed by XAS spectra of SWCN [28-30] and graphite [31]. The similarity between XAS spectra for graphite and SWCN shows clear evidence of the π-conjugated system in SWCN. The hydrogenation of the C-C π-bonds in the walls of SWCN leads to a decrease of the π* resonance intensity as seen in plot (b) of fig. 2. This is accompanied by an increase of intensity between 287 and 290 eV that corresponds to the energy position of the C-H* resonance in hydrocarbons [14]. However, this resonance is not as sharp as seen for well defined molecular species with C-H bonds [14]. The broad nature of the C-H* resonance can be explained by variation of the C-H* resonance energy depending on the local degree of hydrogenation and the diameter of the various nanotubes.

We have thus demonstrated that SWCN films can be hydrogenated to store hydrogen, but two important questions remain. One is to determine the amount of hydrogen stored in SWCN and the other is related to the reversibility of such hydrogenation.

The amount of hydrogen attached to the SWCN was quantified by decomposition of the C1s XPS spectrum of hydrogenated SWCN. In fig. 3, peak (1) and peak (2) are assignable to non-hydrogenated and hydrogenated carbon atoms, respectively. The positions of both peaks (1) and (2) agree well with computed C1s binding energy for unaffected and hydrogenated C atoms in nanotubes. Based on the intensities of peak (2) and peak (1) we evaluate the amount of hydrogenated carbon atoms to $65 \pm 15$ %. We estimate the hydrogen capacity to be $5.1 \pm 1.2$ weight %. XPS spectra measured at different photon energies resulting in different kinetic energies and, thus, different electron escape depths, demonstrate that the hydrogenation takes place inside of the nanotube bundles and not only on their surface.

To check the reversibility of the hydrogenation of SWCN we performed two cycles of hydrogenation/dehydrogenation. The corresponding XPS C1s spectra are shown in fig. 4. Flash annealing up to 600 $^o$C after the first hydrogenation leads to the disappearance of the shoulder in the C1s spectrum (fig. 4, spectrum (c)). Further annealing at higher temperatures did not change the C1s spectral shape so we conclude that hydrogen chemisorbed in SWCN film is released at temperatures below 600 $^o$C. From spectrum (d) we see that the second hydrogenation of the SWCN film leads to a similar change in the C1s spectrum as we observed in spectrum (b). Annealing the sample again to 600 $^o$C restores the C1s spectrum shape similar to spectrum (c) (see fig. 4, spectrum (e)). This means that hydrogenation of the SWCN is possible in a reversible manner. XPS measurements showed that the total amount of carbon in the SWCN sample stayed constant during the hydrogenation/dehydrogenation cycles. From the XPS results we also see that the FWHM of the C1s peak in spectra (c) and (e) is somewhat larger than of the C1s peak in spectrum (a). According to [32] such increase of FWHM can be due to an increase in the number of defects. The increase of the D to G band ratio in the Raman spectrum after two hydrogenation/dehydrogenation cycles (fig. 4, insert) also points to an increase of defects in the SWCN films.

The present results indicate that it is possible to form local C-H bonds by a chemical interaction between hydrogen and SWCN. To fully realize hydrogen storage in SWCN it will be essential to find means to dissociate hydrogen and to fine tune the energetics of the C-H bonds to allow for hydrogen release at 50-100 $^o$C. The former can be solved using an appropriate metal catalyst for hydrogen dissociation and the latter accomplished by using a SWCN with a well defined radius. Theoretical calculations indicate that the C-H bond is weaker for hydrogenated SWCN with larger radius [5,8,33].

In summary, we have investigated the interaction of atomic hydrogen with high quality SWCN films using atom specific spectroscopic probes that are sensitive to C-H bond formation. We found that $65 \pm 15$ at % of the carbon atoms in the walls of SWCN can be hydrogenated to form C-H bonds. This corresponds to a $5.1 \pm 1.2$ weight % of hydrogen capacity. The hydrogenated SWCN are stable at room temperature, and hydrogen is released by heating to 600 $^o$C. This hydrogenation/dehydrogentation process can be cycled.

We acknowledge Donghui Lu at beamline 5-1 at SSRL and Tolek Tyliszczak and Hendrik Bluhm at beamline 11.0.2 at ALS for their technical support. The participation of RD was made possible by the Bavaria California Technology Center (BaCaTeC). This work was supported by Global Climate and Energy Project operated by Stanford University and carried out at the Stanford Synchrotron Radiation Laboratory, a national user facility operated by Stanford University on behalf of the U.S. Department of Energy, Office of Basic Energy Sciences.

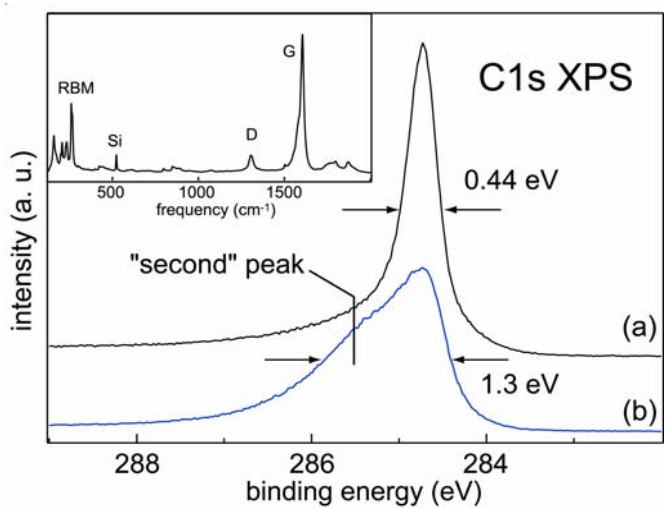

Fig. 1 C1s XPS spectra of the clean SWCN film (a) and SWCN film after hydrogenation (b). Spectra are normalized to the area of the C1s peak. Insert: Raman spectrum of clean SWCN film.

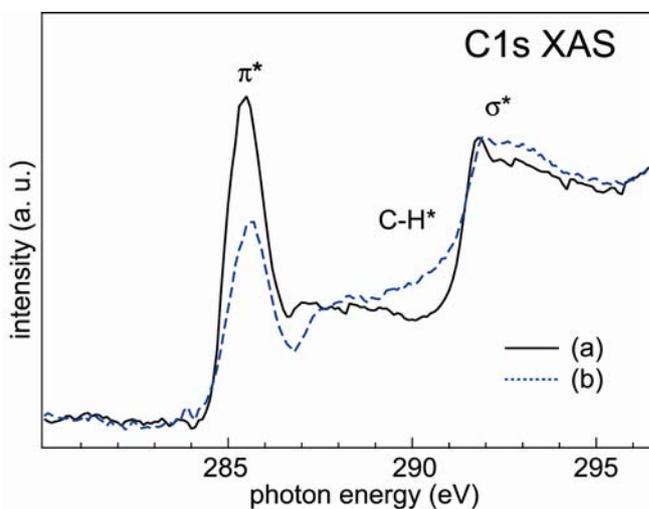

Fig. 2 Carbon K-edge XAS spectra of clean SWCN film (a) and SWCN film after hydrogenation (b).

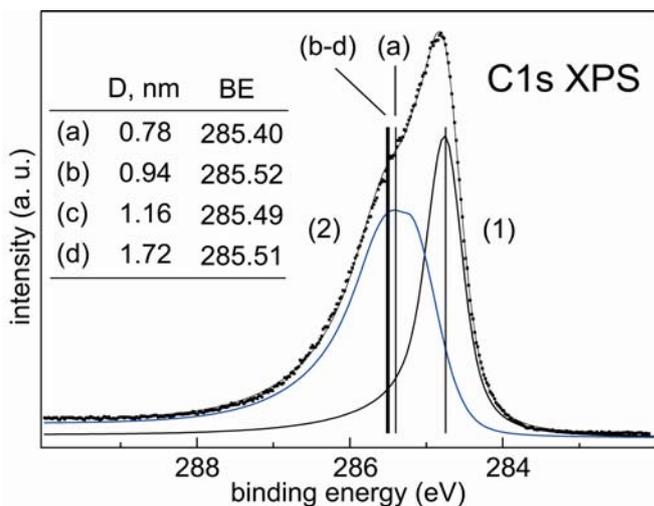

Fig. 3 Decomposition C1s XPS spectrum (plot (b), fig. 1) into peak (1) (unaffected carbon atoms), and peak (2) (hydrogenated carbon atoms). The insert indicates the calculated C1s binding energies (BE) for hydrogenated nanotubes in different diameters (D) which are shown as vertical lines.

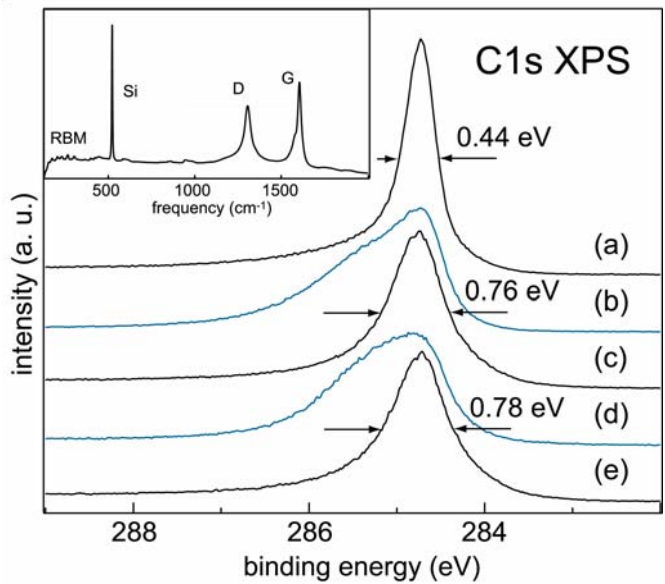

Fig. 4 C1s XPS spectra of SWCN film exposed to the two cycles of hydrogenation/dehydrogenation: (a) clean SWCN film, (b) hydrogenated SWCN film, (c) SWCN annealed at 600 °C, (d) hydrogenated SWCN film, (e) SWCN annealed at 600 °C. Spectra are normalized to the area of the C1s peak. Insert: Raman spectrum of SWCN film after two cycles of hydrogenation/dehydrogenation.